\documentclass[aps,pra,floatfix,twocolumn,superscriptaddress,showpacs]{revtex4}

\pdfoutput=1
\usepackage{graphicx}
\usepackage{amssymb}
\usepackage{amsmath}

\newcommand {\ket}[1] {|#1 \rangle}

\newcommand{\lr}[1]{\left( #1 \right)}

\newcommand{\mean}[1]{\langle #1 \rangle}

\newcommand{\tr}[1]{\textrm{tr}\left\{#1\right\}}

\begin{document}

\title{Measuring entanglement growth in quench dynamics of bosons in an optical lattice}

\author {A.~J.~Daley}
\affiliation{Department of Physics and Astronomy, University of Pittsburgh, Pittsburgh, Pennsylvania 15260, USA}

\author{H.~Pichler}
  \affiliation{Department of Physics and Astronomy, University of Pittsburgh, Pittsburgh, Pennsylvania 15260, USA}
  \affiliation{Institute for Theoretical Physics,
  University of Innsbruck, A-6020 Innsbruck, Austria}
  \affiliation{Institute
  for Quantum Optics and Quantum Information of the Austrian Academy
  of Sciences, A-6020 Innsbruck, Austria}

\author{J.~Schachenmayer}
\affiliation{Department of Physics and Astronomy, University of Pittsburgh, Pittsburgh, Pennsylvania 15260, USA}
\affiliation{Institute for Theoretical Physics,
  University of Innsbruck, A-6020 Innsbruck, Austria}\affiliation{Institute
  for Quantum Optics and Quantum Information of the Austrian Academy
  of Sciences, A-6020 Innsbruck, Austria}

\author{P.~Zoller}
\affiliation{Institute for Theoretical Physics,
  University of Innsbruck, A-6020 Innsbruck, Austria}
  \affiliation{Institute
  for Quantum Optics and Quantum Information of the Austrian Academy
  of Sciences, A-6020 Innsbruck, Austria}
\affiliation{Department of Physics, Harvard University, Cambridge, MA 02138, USA}
\affiliation{Joint Quantum Institute, National Institute of Standards and Technology, and University of Maryland, Gaithersburg, Maryland 20899, USA}

 \date{May 7, 2012}

\pacs{37.10.Jk, 03.65.Ud, 03.67.Mn, }

\begin{abstract} We discuss a scheme to measure the many-body entanglement growth during quench dynamics with bosonic atoms in optical lattices. By making use of a 1D or 2D setup in which two copies of the same state are prepared, we show how arbitrary order R\'enyi entropies can be extracted using tunnel-coupling between the copies and measurement of the parity of on-site occupation numbers, as has been performed in recent experiments. We illustrate these ideas for a Superfluid-Mott insulator quench in the Bose-Hubbard model, and also for hard-core bosons, and show that the scheme is robust against imperfections in the measurements.
\end{abstract}

\maketitle

Entanglement is a basic feature of many body quantum systems \cite{entanglement}, and underlies the complexity of simulating quantum physics on a classical computer \cite{Vidal03}. The exponential scaling of resources to represent and propagate a general many body quantum state on a classical device has motivated the development of quantum simulators \cite{simulator}, and significant progress has been made in building both analog and digital quantum simulators with cold atoms and ions for equilibrium and non-equilibrium dynamics. This is exemplified by quantitative measurement of phase diagrams, studies of quantum phase transitions, and quench dynamics. An outstanding challenge, however, is direct measurement of (potentially large scale) entanglement, and monitoring entanglement growth in non-equilibrium dynamics.  Below we address these questions by discussing measurement scenarios for entanglement entropies, using multiple copies of a quantum system and measurements with a quantum gas microscope \cite{Bak09,She10}. We illustrate these ideas in the context of quench dynamics of bosons in 1D optical lattices. This example is motivated by recent experiments \cite{Tro12}, where quench dynamics were observable for times not accessible to (classical) t-DMRG simulations of Hubbard dynamics \cite{Vidal04,tdmrg1,tdmrg2,dmrgrev} due to entanglement growth \cite{Sch08,Sch08b,Pro07,Ven09}. Here the measurement protocol will directly reveal this {\em entanglement growth}, and simultaneously monitor the {\em purity of the total system state}. By comparing copies, these tools will also provide a protocol for the {\em verification of a quantum simulator}. 

We are interested in quantum dynamics of an (ideally) isolated quantum system as represented by our atomic quantum simulator. In particular, we study a system where we prepare an initial state $|\Psi(0)\rangle$, which evolves with a Hamiltonian $H$ as $|\Psi(t)\rangle=\exp(-iHt/\hbar) |\Psi(0)\rangle$. If the system can be divided into two subsystems $A$ and $B$ and is in a pure state at time $t$, $\rho=|\Psi \rangle \langle \Psi |$, then the entanglement of the system can be characterized in terms of the entropy of the reduced density matrix, $\rho_A=\textrm{tr}_B\{\rho\}$. This is commonly computed as the von Neumann entropy $S_{VN}(\rho)=-\textrm{tr}\{\rho \log \rho\}$. If $A$ and $B$ are in a product state $|\Psi\rangle=|\Psi_A\rangle\otimes |\Psi_B\rangle$, then $\rho_A$ will also represent a pure state with $\textrm{tr}\{\rho_A^2\}=1$, and $S_{VN}(\rho_A)=0$.
Below we will discuss measurement of the R\'enyi entropy of order $n\geq 2$, $S_n(\rho)=\frac{1}{1-n}\log \textrm{tr}\{\rho^n\}$, which gives quantitative bounds for a variety of measures of entanglement, e.g., the concurrence \cite{Min05,Aol06,Min07,Min07_2}, and also $S_{VN}(\rho)$. For example, we know that $S_{VN}(\rho)\,\,[=\lim\limits_{n\rightarrow 1}S_{n}(\rho)]$, and as $dS_n(\rho)/dn\leq 0$, $S_{VN}(\rho)\geq S_2(\rho)$. From $d^2S_{n}/dn^2\geq 0$ a stronger bound $S_{VN}(\rho)\geq 2S_2(\rho)-S_3(\rho)$ can be obtained \cite{Zyc03} if we measure multiple R\'enyi entropies.

\begin{figure}[tb]
\begin{center}
\includegraphics[width=8cm]{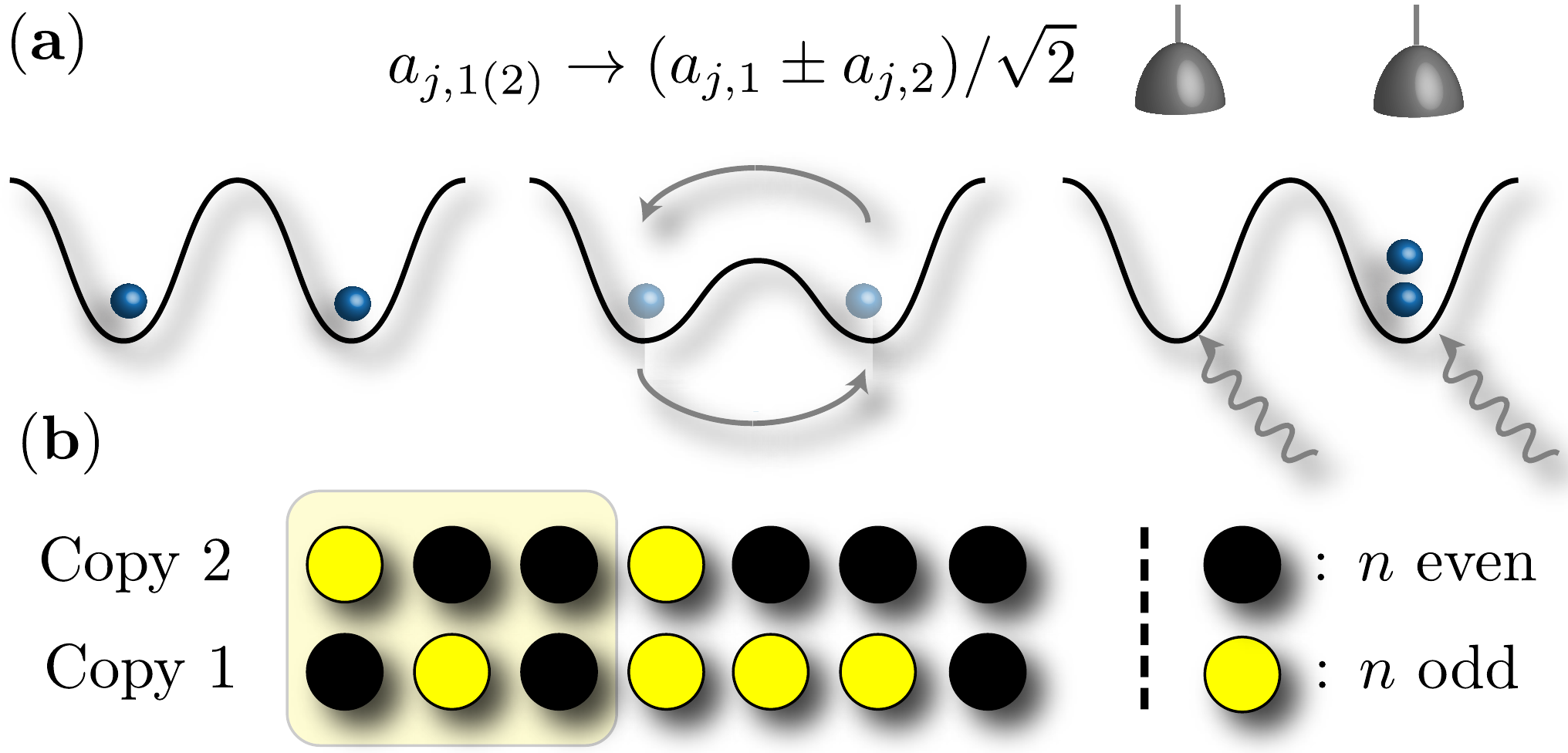}
\caption{(a) Measurement of $n=2$ R\'enyi entropy for bosons in an optical lattice. First, two instances of the many-body state are produced (shown here for a single site in a 1D chain or 2D plane). Then tunnelling is switched off within each copy, and the barrier between the copies lowered to realize a beamsplitter operation between the copies. Finally, the parity of the atom number is measured at each site. This measurement is repeated to obtain expectation values for the swap operator $V_2$, from which the R\'enyi entropy can be computed (see text). (b) Example measurement outcome for a single shot on a quantum chain. Here the measurement result for the whole system swap operator $V_2^{\{1,\dots,7\}}$ is $1$, since the total number of particles in Copy 2 is even. For the swap of the first three sites $V_2^{\{1,2,3\}}$ is $-1$, since this number is odd.}
\end{center}
\end{figure}

In order to measure the R\'enyi entropy $S_n(\rho)$ for a state $\rho$, we require $n$ copies of the state prepared in parallel, and the possibility to implement operations that exchange the $n$ copies \cite{Eke02,Hor02,Mou04,Pal05,Bre03,Min05}. As shown in Ref.~\cite{Eke02}, the quantity $\textrm{tr}\{\rho^{n}\}$ can be written \cite{Eke02,suppl} as $\textrm{tr}\{\rho^{n}\}=\textrm{tr}\{V_n\rho^{\otimes n}\}$, where the shift operator on $n$ copies, $V_n\ket{\psi_1}\dots\ket{\psi_n}=\ket{\psi_n}\ket{\psi_1}\dots\ket{\psi_{n-1}}$. Therefore, the measurement of the R\'enyi entropy can be reduced to determining the expectation value $\langle V_n \rangle$ on the $n$ copies. Measurements of inner products can be made in this way by entangling a state with auxilliary qubits or a quantum switch \cite{Eke02, Mue09, Aba12, Mic04}, as has been demonstrated for a few entangled photons \cite{Bov05,Wal07,Schm08}. In Ref.~\cite{Mou04} it was shown that a beamsplitter operation on two copies is sufficient to measure the purity $\mathrm{tr}\{\rho^2\}$ for bosonic systems. Our study of entanglement growth is based on generalizing these techniques to measure R\'enyi entropies of arbitrary order $n$. Remarkably, the corresponding experimental tools (controlled tunnelling between multiple copies and measurement of onsite occupation numbers modulo $n$) are now available with single-site addressing in a quantum gas microscope \cite{Bak09,She10}.

Below we discuss the protocol to measure $S_n(\rho)$ for arbitrary $n$, and summarize the details of the procedure for the simplest example $n=2$.
We also provide a simplified scheme for hard-core bosons, and discuss the robustness of measurements with respect to imperfections. We note throughout that this scheme allows simultaneous measurement of $S_n(\rho)$ for the whole state and for every reduced subsystem (i.e., $\rho=\rho_A$) \footnote{Note that this is useful in determining the scaling of entanglement with the size of reduced blocks}. Thus, while measurement of the entanglement in this form is based on the assumptions (i) that the initial states for the whole system are pure, and (ii) that each evolves under the same Hamiltonian, these assumptions can be checked directly by measuring $S_n(\rho)$ for the whole system. In quench dynamics, these assumptions should be well fulfilled in experiments beginning from low-entropy states \cite{Sim11, Rabl03}. Moreover, by monitoring the copies over time and measuring, e.g., $\mathrm{tr}\{\rho_1 \rho_2\}$ \cite{Eke02,suppl}, this scheme provides a means to verify a quantum simulator, determining whether the evolution of the copies is coherent and identical on the level of many-body wavefunctions.

\emph{Scheme for arbitrary $n$}--- The procedure to measure the R\'enyi entropy $S_n(\rho)$ consists of three steps: \textbf{(1)} $n$ identical instances of the many-body state are prepared in parallel (either in $n$ 1D chains, or in $n$ planes in 2D). This can be performed, e.g., by beginning from a low entropy initial state such as a Mott insulator \cite{Sim11, Rabl03}, and manipulating the lattice potential identically for the two copies, i.e., allowing them to evolve under the same Hamiltonian. In this step, the lattice depth between the copies must remain large so that these are isolated from each other. \textbf{(2)} We then make the lattice deep within each copy of the state, to prevent tunnelling, and perform a discrete Fourier transform $U_{n}^{FT}$ operation on the copies. If the bosonic annihilation operator for site $i$ in copy $c\in \{1,\ldots, n\}$ is $a_{i,c}$, then
\begin{align}
U_n^{FT}: a_{j,k}\rightarrow \frac{1}{\sqrt{n}}\sum_{l=1}^n a_{j,l}\mathrm{e}^{\mathrm{i}\frac{2\pi}{n}(k-1)(l-1)}.
\end{align}
This can be achieved by a successive application of tunnelling between the copies, and shifting the relative potential depths (to produce elements analogous to beam-splitters and phase-shifters in the optical implementation of this operation \cite{Rec94}). This operation is very simple for small $n$, as discussed below, and should be performed on the whole system in parallel. \textbf{(3)} We then perform a site-resolved measurement of the onsite particle number $n_{i,c}$ in each copy, modulo $n$. We can then determine the measured value of the swap operator $V_{n}^{\mathcal{R}}$ for all possible subsystems being swapped, $\mathcal{R}$, in parallel (where we note that $\mathcal{R}$ can also denote the whole system), as these commute. The possible measurement outcomes for the swap operations $V_n^{\mathcal{R}}$, $\{\mathrm{e}^{\mathrm{i}\frac{2\pi}{n}j}|j=1\dots n\}$ can then be computed from the number measurements as $\prod_{j\in\mathcal{R}}\mathrm{e}^{\mathrm{i}\frac{2\pi}{n}\sum_{c=1}^{n}n_{j,c}(c-1)}$. We illustrate this scheme for $n=2$ below, and $n=3$ in the supplementary material. 

\emph{Scheme for $n=2$}--- As an example, the measurement of the R\'enyi entropy for $n=2$ \cite{Mou04} is illustrated in Fig.1. In step (2), the FT for two copies is  a \textit{beamsplitter operation}, 
\begin{equation}\label{bs}
a_{i,1}\rightarrow (a_{i,1}+a_{i,2})/\sqrt{2},\quad
a_{i,2}\rightarrow (a_{i,2}-a_{i,1})/\sqrt{2}.
 \end{equation}
This can be achieved by lowering the barrier between the two copies (e.g., using a superlattice \cite{Foe07}) and allowing the atoms to tunnel from one site to its copy for a time $T=\pi/(4 J_{12})$ where $J_{12}$ is the tunnelling rate. For this step to work in this form, interactions between atoms have to be turned off during this process, e.g. via a Feshbach resonance. We will show below that this requirement can be relaxed in an alternative method for hard-core bosons. The number measurement modulo $2$ in step (3) is then a parity measurement of the site-resolved occupation number in each site and copy $\{n_{i,c}\}$, exactly as is performed in recent quantum gas microscope experiments \cite{Bak09,She10}. 

The measured value of the swap operator, $V_{2}^{\mathcal{R}}$, can be computed as $(-1)^{\sum_{i\in \mathcal{R}} n_{i,2}}$, i.e., \textit{simply by determining whether the total atom number in block $\mathcal{R}$ of copy 2 after the measurement is even or odd}. This process is illustrated in Fig.~1b for one measurement instance, and repeating this process allows the expectation value $\langle V_2^{\mathcal{R}}\rangle$ to be computed. 
To obtain this relationship between the occupation numbers after the beamsplitter operation and $\langle V_2^{\mathcal{R}}\rangle$, we note that each possible measurement outcome $\{n_{i,c}\}$ corresponds to the measurement of one of the two eigenvalues of the (hermitean) operators $V_2^{\mathcal{R}}$, which are $\pm1$. This can be seen as follows: The eigenspaces of $V_2^{\mathcal{R}}$ are the subspaces of the total Hilbert space that are (anti-)symmetric with respect to exchange of the two copies, which we denote by $\mathcal{H}_{\mathcal{R}}^{+}$ ($\mathcal{H}_{\mathcal{R}}^{-}$). Since the swap on a subsystem $\mathcal{R}$ can be constructed by local swaps of the sites $j\in \mathcal{R}$, $V^{\{j\}}$, we have $V_2^{\mathcal{R}}=\prod_{j\in \mathcal R} V_2^{\{j\}}$. Since all of the $V_2^{\{j\}}$ commute ([$V_2^{\{j\}},V_2^{\{k\}}]=0$) we need to consider only a single site. Now, if we denote the annihilation operator for bosons on site $i$ of copy $c$ by $a_{i,c}$, $c\in\{1,2\}$, the tunnelling procedure in step (2) gives us Eq.~(\ref{bs}).
This maps the symmetric subspace of the two modes $a_{j,1}$ and $a_{j,2}$ to the subspace of states with an \textit{even} number of atoms in mode $a_{j,2}$. The anti-symmetric subspace is mapped to states with an \textit{odd} number of bosons in mode $a_{j,2}$. This can be seen by noting that the anti-symmetric subspace $\mathcal{H}_j^{-}$ of the two modes at site $j$ is spanned by the states $\{(a_{j,1}^{\dag}-a_{j,2}^{\dag})^{2n+1}\ket{\rm{vac}}\}$, while the symmetric one $\mathcal{H}_j^{+}$ is spanned by $\{(a_{j,1}^{\dag}-a_{j,2}^{\dag})^{2n}\ket{\textrm{vac}},(a_{j,1}^{\dag}+a_{j,2}^{\dag})^n\ket{\rm vac}\}$ (where $n=0,1,2,\dots$). Under the above operation the first set is mapped onto $\{(a_{j,2}^{\dag})^{2n+1}\ket{\rm{vac}}\}$, i.e. states with an \textit{odd} number of atoms in $a_{j,2}$, while the second is mapped onto $\{(a_{j,2}^{\dag})^{2n}\ket{\textrm{vac}},(a_{j,1}^{\dag})^n\ket{\rm vac}\}$ with an \textit{even} number of atoms in $a_{j,2}$. Therefore, the measurement outcome of $V^{\{j\}}$ is $+1$ if $n_{j,2}$ is even and $-1$ if $n_{j,2}$ is odd.

\begin{figure}[tb]
\begin{center}
\includegraphics[width=0.49\textwidth]{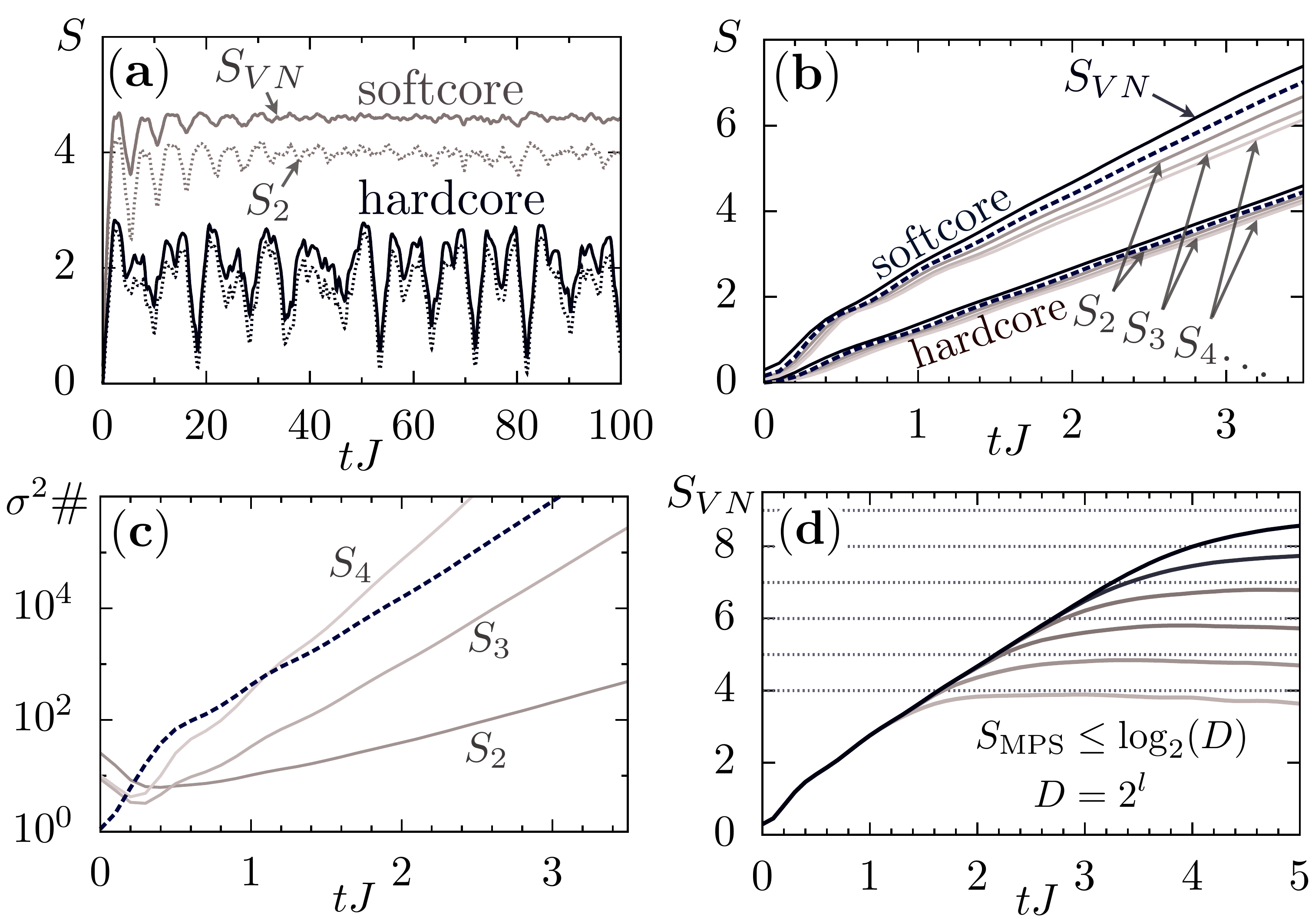}
\caption{Entanglement buildup and measurement, both for (i) \emph{softcore} bosons in 1D after a quench from a Mott Insulator at $U/J=10$ to a superfluid at $U/J=1$; and (ii) \emph{hardcore} bosons with tunnelling $J$, beginning from an initial state where every second lattice site is occupied. (a) Quenches in small systems with $8$ particles on $8$ lattice sites, showing Von Neumann and R\'enyi entropies for a bipartite splitting in the center of the system; (b) the same as (a), but for a system of $30$ particles on $30$ sites, and also including $2S_2-S_3$ (dotted line); (c) The number of
single shot measurements ($\#$) required to determine $S_n$ for the Bose-Hubbard quench in (b) with a relative accuracy $\sigma$. For $2S_2-S_3$, measurements are distributed between ($S_2$) and ($S_3$) to minimize the total number. (d) Limitations of t-DMRG simulations for the Bose-Hubbard quench based on different bond dimensions $D$ \protect\cite{dmrgrev}, shown here with $D=2^l$ for $4\leq l \leq 9$, with corresponding bound $S_{VN}=l$..}

\label{fig:fig:entanglement_creation}
\end{center}
\end{figure}

\emph{Example of a Mott Insulator-Superfluid Quench}--- We now illustrate the entropy measurement for a quench in the Bose-Hubbard model, where quantum dynamics generates substantial entanglement in a short time \cite{Kol07,Lau08,Una10}. In Fig.~2 we plot results obtained via t-DMRG calculations from dynamics (in each copy of the system) described by the Bose-Hubbard model ($\hbar\equiv 1$), $H_{BH}=- J \sum_{\langle i,j \rangle} a^\dag_i a_j +(U/2) \sum_i a^{\dag 2}_i a_i^{\vphantom{\dag}{2}}$, where $J$ is the tunnelling rate between neighbouring sites $\langle \ldots \rangle$, and $U$ is the onsite interaction strength. We plot both (i) a quench for \emph{softcore} bosons from a Mott Insulator state with $U/J=10$ to a superfluid (SF) at $U/J=1$,
and (ii) a quench for \emph{hardcore} bosons $U\rightarrow \infty$, where we begin from an initial state with one particle on every second lattice site. Such a state could be produced in a superlattice potential \cite{Tro12}, or using recently demonstrated techniques to directly remove atoms from an initial Mott Insulator state \cite{Wei11}.

Fig.~2a shows $S_{VN}$ and $S_2$ calculated for a bipartite splitting in the center of a small system with $8$ particles on $8$ lattice sites. In this size of system, we see that the entanglement entropy saturates as the system thermalizes for softcore bosons \cite{Rig08}. In the hardcore case, the system is integrable, and we see large oscillations in the entanglement entropies, instead of thermalization \cite{Rig09,Kin06}. Fig.~2b shows a comparison between the von Neumann entropy, $S_{VN}$, and R\'enyi entropies $S_n$ (for $n=2,3,4$), after a quench in a larger system with $30$ particles on $30$ sites. We see rapid growth in the entanglement of two halves of the system, and we note that the R\'enyi entropies provide relatively good bounds for the Von Neumann entropy, especially if we use the the bound $S_{VN} \geq 2S_2-S_3$, which is shown as the dashed line in the figure. Fig.~2b shows the number of
single shot measurements required to determine the quantities in Fig.~2b with a relative error of
$\sigma$.  We note that in these larger systems, the t-DMRG simulations we use to compute dynamics are limited to simulating short
timescales, since the von Neumann entropy increases linearly with time \cite{Sch08,Sch08b}. A matrix product state (MPS) with a bond dimension $D$ \cite{dmrgrev} is capable of representing a maximum von Neumann entropy up to
$\log_2(D)$. In Fig.~2c, we show the evolution of the entropy as a function of time for $D=2^l$ with $4\leq l \leq 9$, showing clearly the different bounds for the entanglement that can be represented. In the case of $D=512$ we can faithfully simulate time-scales only up to $tJ \sim 3$. The simulation times scale $\sim D^3$, and in an experiment, substantially higher entanglement entropies could be generated than are accessible in reasonable time on a classical computer. This could be demonstrated directly by measuring $S_n(\rho)$ in an experiment.

\emph{Simplified scheme for hard-core bosons}---
The measurement scheme presented above relies on the ability to turn off interactions between the atoms in order to realize the beamsplitter operations, which can be challenging for some atomic species. However, in certain cases this requirement can be relaxed, e.g., in the case of hardcore bosons $U\gg J$, where we have at most one atom per site, the measurement can be performed without switching the interaction strength. For measurement of $S_2(\rho)$, the symmetric subspace at site $j$, $\mathcal{H}_j^{+}$, is spanned by $\{\ket{\textrm{vac}},(a_{j,1}^{\dag}+a_{j,2}^{\dag})\ket{\textrm{vac}},a_{j,1}^{\dag}a_{j,2}^{\dag}\ket{\textrm{vac}}\}$ while the antisymmetric one, $\mathcal{H}_j^{-}$, is spanned by $(a_{j,1}^{\dag}-a_{j,2}^{\dag})\ket{\textrm{vac}}$. Of those four states the tunnel coupling in step (2) of the measurement scheme only affects states with one particle in total on the copies of the lattice site. The other two states are invariant (either because there are no atoms or because tunneling is suppressed due to the hardcore constraint). Thus we map $\mathcal{H}_j^{+}\rightarrow \{\ket{\textrm{vac}},a_{j,1}^{\dag}\ket{\textrm{vac}},a_{j,1}^{\dag}a_{j,2}^{\dag}\ket{\textrm{vac}}\}$ and $\mathcal{H}_j^{-}\rightarrow a_{j,2}^{\dag}\ket{\textrm{vac}}$ in step (2). Measurement of the onsite atom number can then be used to directly distinguish between the two eigenspaces of $V_2^{\{j\}}$, and thus determine the measurement result. We extend this to the measurement of $S_3(\rho)$ in the supplementary material. 

\begin{figure}[tb]
\begin{center}
\includegraphics[width=0.49\textwidth]{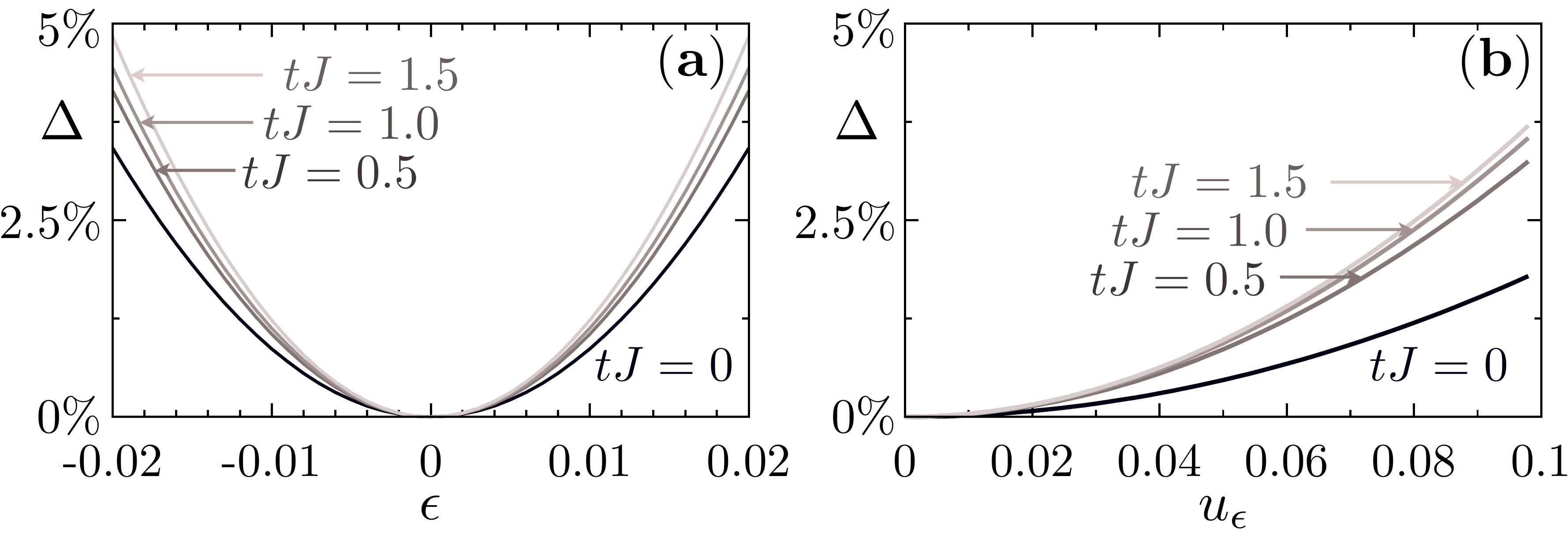}
\caption{
Errors $\Delta=\left({{\rm tr} (\rho^2_{a\dots M})-\langle
\prod_{j=a}^M e^{i\pi\hat n_j}\rangle_{t}}\right) /{{\rm tr}
(\rho^2_{a\dots M})}$ that are introduced by an imperfect
beamsplitter operation, determined via t-DMRG simulation of the Bose-Hubbard quench in Fig.~2a and resulting measurement process. (a) the effect of timing errors, with $T=\pi/(4 J_{12}) + \epsilon$; (b)
error introduced by a finite interaction strength
$u_\epsilon=U_\epsilon/J_{12}$ during the measurement.}
\end{center}
\end{figure}

\begin{figure}[tb]
\begin{center}
\includegraphics[width=0.49\textwidth]{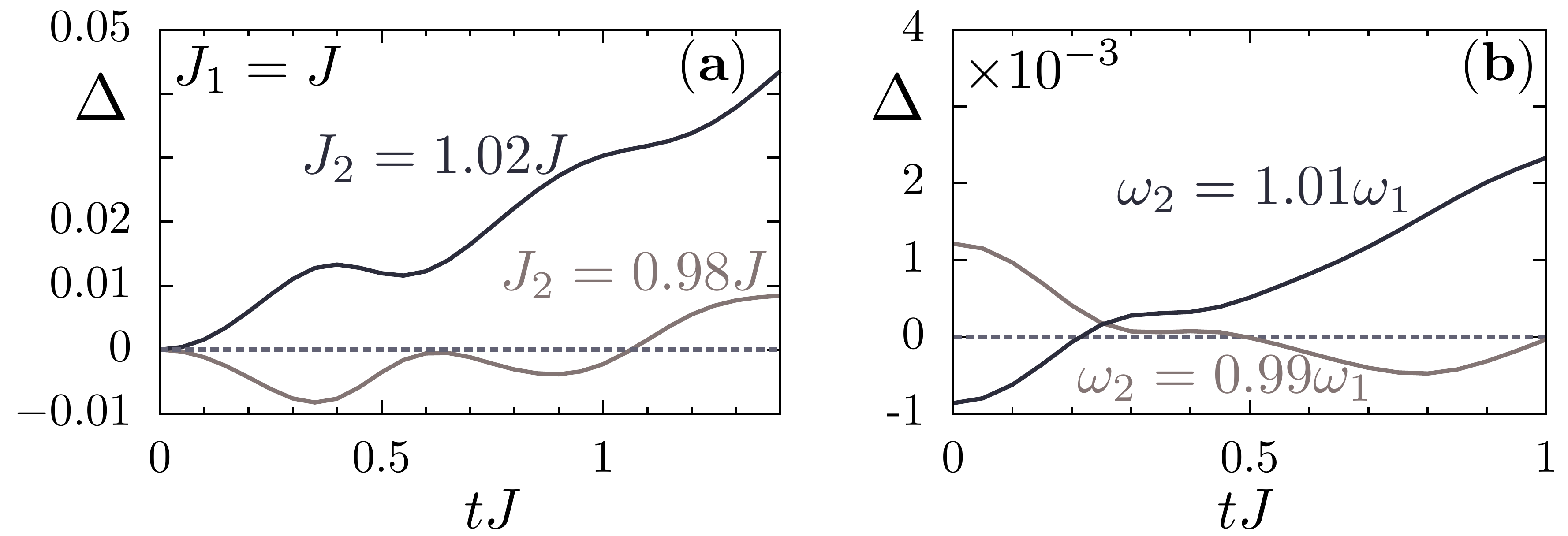}

\caption{
Deviations resulting from slightly different evolutions in the two copies, $\Delta=\left({\rm tr} (\rho^2_{a\dots M})-\langle
\prod_{j=a}^M e^{i\pi\hat n_j} \rangle_{t}\right) /{{\rm tr}
(\rho^2_{a\dots M})}$. (a) Results for slightly different tunneling amplitudes in
each copy after the MI-SF quench from Fig.~2a. (b) Results for a case where each copy has an additional harmonic trapping
potential with $\omega_1=2\times 10 J (N/2)^{-2}$ for the first copy, and we otherwise perform the same quench as in (a).}
\end{center}
\end{figure}

\emph{Robustness against imperfections}---We now analyze the robustness of the measurements to errors in the measurement steps, especially imperfect implementation of the beamsplitter operations in step (2). Two main imperfections can occur here, arising from timing errors or residual interactions. In the case of timing errors where $T=\pi/(4 J_{12}) + \epsilon$, we can show analytically \cite{suppl} that the measured value of entanglement will always be smaller than the actual entanglement by an amount proportional to $\epsilon^2$, thus leaving a clear lower bound. In Fig.~3a we plot results obtained from a full t-DMRG simulation of the Bose Hubbard quench from Fig.~2, and the subsequent measurement operation. For timing errors on the order of $1\%$ we find a $\Delta$ on the order of $1\%$
which increases slowly with increasing entanglement. In the same calculation, we also consider errors introduced by residual interparticle interactions $U_\epsilon/J_{12}$ present during the measurement operation. From Fig.~3b we see that even
in the case of an interaction of $10\%$ of the beam-splitter tunneling
amplitude $J_{12}$ we find resulting errors on the order of only $1\%$
which increase slowly with time.

\emph{Non-identical copies and verification of a quantum simulator}---Another type of imperfection, is where the ``copies'' undergo different dynamics before the measurement process, e.g., due to different Hamiltonians. Such errors are relevant for characterising the accuracy of the quantum simulator itself, as well as constituting an error in the measurement scheme. In Fig.~4a we show results when we have slightly different tunneling amplitudes in
each copy after the Bose-Hubbard quench. We find that the error introduced by
$2\%$ deviations in $J$ correspond to a $\Delta$ of $1\%$ on a
time-scale of $tJ \sim 1$, which increases with time. Fig.~4b shows results from a trapped case, where we begin with slightly different trapping potentials in the individual copies, and then perform the same interaction quench from $U/J=10$ to $U/J=1$. On a
time-scale $tJ\sim 1$ errors of $1\%$ lead to deviations on the order
of $0.1\%$.  In this case, the quantity that is actually measured is $\mathrm{tr}\{\rho_1 \rho_2\}$, where $\rho_1$ and $\rho_2$ are the density operators of the different copies. This can be used as a tool to verify the quantum simulator in the case that the evolutions should be identical, or to investigate dynamics with the evolution Hamiltonians being set to slightly different values. The latter could be interesting in regimes of the Bose-Hubbard model expected to display signatures of quantum chaos \cite{Cas09,Kol04,Ven09}, where the inner product might decay exponentially in analogy with single-particle systems exhibiting quantum chaos \cite{Gar97,Per91,Gor06}.

\emph{Summary}---We have analyzed measurement of R\'enyi entropies of 1D bosons in an optical lattice during a quench, using techniques currently available in quantum gas microscope experiments. Such a tool can be used to verify the coherence and accuracy of a 1D and 2D quantum simulator, and determine the entanglement growth in a quantum quench, helping the experiments to diagnose regimes where dynamics can be realized that are not accessible to present classical computations. 

{\em Acknowledgments:} We thank I.~Bloch, M.~Greiner, M.~Lukin, A. L\"auchli, F. Mintert, E.~Rico Ortega, T.~Pichler and M. Tiersch, for helpful and motivating discussions. In the final stages of this work we have become aware of Ref.~\cite{Aba12} on measuring entanglement entropies of Hamiltonian ground states using a single spin as a quantum switch connecting $n$ copies of quantum systems. We thank D. Abanin and E. Demler for discussions comparing our approaches. PZ and HP thank the Harvard Physics Department and the Joint Quantum Institute, University of Maryland, and HP thanks the University of Pittsburgh for hospitality. 
Work supported in part by the Austrian Science Fund through SFB F40 FOQUS, and by a grant from the US Army Research Office with funding from the DARPA OLE program. Computational resources  
provided by the Center for Simulation and Modeling at the University of Pittsburgh.

¥

\newpage
\section*{Supplementary material}

\section{Relationship between the R\'enyi entropy and the expectation value of the swap operator}

Let us assume that we have two states $\rho_1$ and $\rho_2$ in separated subsystems $1$ and $2$ with identical Hilbert spaces (these represent the two copies in our measurement scheme). We then define a swap gate $V_2$ on these two states, so that for any two states of the systems $|\psi_{1}\rangle_1$ and $|\psi_{2}\rangle_2$,
\[ V_2|\psi_{1}\rangle_{1}\otimes |\psi_{2}\rangle_{2}=|\psi_{2}\rangle_{1}\otimes|\psi_{1}\rangle_{2}.\]
We then decompose the density operators of the subsystems as $\rho_{1}=\sum_{ij}\rho_{ij}^{(1)}|i\rangle\langle j|$,
$\rho_{2}=\sum_{kl}\rho_{kl}^{(2)}|k\rangle\langle l|$ in the corresponding basis states for each subsystem. We then see that
\begin{eqnarray*}
 \mathrm{tr}\{V_2\rho_{(1)}\otimes\rho_{(2)}\} & = & \mathrm{tr}\left\{ V_2\sum_{ijkl}\rho_{ij}^{(1)}\rho_{kl}^{(2)}|i\rangle\langle j|\otimes|k\rangle\langle l|\right\} \\
 & = & \mathrm{tr}\left\{ \sum_{ijkl}\rho_{ij}^{(1)}\rho_{kl}^{(2)}|k\rangle\langle j|\otimes|i\rangle\langle l|\right\} \\
 & = & \sum_{ijkl}\rho_{ij}^{(1)}\rho_{kl}^{(2)}\delta_{kj}\delta_{il}\\
 & = & \sum_{ik}\rho_{ik}^{(1)}\rho_{ki}^{(2)}=\mathrm{tr}\{\rho_{1}\rho_{2}\}.\end{eqnarray*}
 If $\rho=\rho_1=\rho_2$ (i.e., we have two identical copies), this results in $\textrm{tr}\{V_2\rho\otimes \rho\}=\mathrm{tr}\{\rho^2\}$.
 This result can be readily generalized to the case of permuation of $n$ copies, to obtain $\textrm{tr}\{\rho^{n}\}=\textrm{tr}\{V_n\rho^{\otimes n}\}$, where the shift operator on $n$ copies, $V_n\ket{\psi_1}\dots\ket{\psi_n}=\ket{\psi_n}\ket{\psi_1}\dots\ket{\psi_{n-1}}$.

\section{Required number of measurements to determine $S_n(\rho)$ with prescribed accuracy}
The number of measurements needed to determine the expectation value of $V_n^{\mathcal{R}}$ with a given accuracy can be determined by calculating its variance, which can be calculated in terms of $\mean{V_n^{\mathcal{R}}}$ by noting that $V_n^{\mathcal{R}}V_n^{\mathcal{R}\dag}=1$. 
For $n>2$ the shift operator $V_n$ is no longer hermitian. Therefore  in principle it is not guaranteed that its expectation value is real. However, for our assumption, that we have a state of $n$ copies ($\rho^{n}$), it is. Assuming that this requirement is fulfilled, we need to determine only the real part of its expectation value. (In other words, measure the expectation value of $(V_n+V_n^{\dag})/2$). This reduces the number of measurements, by roughly a factor of $1/2$. (However, if one determines also the imaginary part of $\mean{V_n}$, this assumption can be tested.) 
The number of measurements needed to determine $\mean{\tilde V_n^{\mathcal{R}}}$ with a relative error $\sigma$ is then given by $\#_n^{\mathcal{R}}=(1/2-\mean{V_n^{\mathcal{R}}}^2/2)/(\sigma\mean{V_n^{\mathcal{R}}})^2$ if $n$ is odd, and $\#_n^{\mathcal{R}}=(1/2+\mean{V_{n/2}^{\mathcal{R}}}^2/2-\mean{V_n^{\mathcal{R}}}^2)/(\sigma\mean{V_n^{\mathcal{R}}})^2$ if $n$ is even. This shows that the effort to determine R\'enyi entropies of higher orders grows exponentially with $n$ (Note that one has $\textrm{tr}\{\rho^m\}\geqslant (\textrm{tr}\{\rho^n\})^{m/n}$ if $n>m$). However, measurement of $S_n(\rho)$ for several low values of $n$ can already be useful in finding bounds for the von Neumann entropy $S_{VN}$.

\section{Measurement of the $n=3$ R\'enyi entropy}

Here we give the specific example for implementation of the quantum Fourier transform (FT) for measurement of the $n=3$ R\'enyi entropy.
The unitary map associated with the FT is generated in three steps, that involve tunnelling between the copies, and shifting the relative potential depths between copy $\mu$ and $\nu$: $T_{\mu,\nu}(\alpha,\phi): a_{\mu}\rightarrow a_{\mu}\cos(\alpha)-a_{\nu}\sin(\alpha); a_{\nu}\rightarrow (a_{\nu}\cos(\alpha)+a_{\mu}\sin(\alpha))e^{-i\phi};$. One has (up to local phases) $U_3^{FT}=T_{2,1}(\pi/4, \pi/2)T_{3,1}(\arctan(1/\sqrt{2}), 2 \pi/3)T_{3,2}(-\pi/4, \pi/3)$. Note that the second step, the ``beam splitter'' between copy 1 and 3 (which are not neighboring) can be broken down to three beam splitting operations that involve only neighboring copies.\\
For arbitrary $n$ the FT can be achieved with $n(n-1)/2$ beam splitter operations \cite{Rec94}. If only beam splitters between neighboring sites are allowed one can get it with $ (n-1) n (2 n-1)/6$ of these operations.\\

\section{Simplified measurement of $S_3(\rho)$ for hardcore bosons}

Here we show that we can also measure $\tr{\rho^3}$ for hardcore bosons without switching off the interparticle interactions. To do so, we apply the same sequence of lattice modulations that would produce the $U_3^{FT}$ in the absence of interactions. The states $\ket{\textrm{vac}}$ and $a_{j,1}^{\dag}a_{j,2}^{\dag}a_{j,3}^{\dag}\ket{\textrm{vac}}$ (correspondingto the eigenspace of $V_3^{\{j\}}$ with eigenvalue one), are unaffected by this. The eigenstates states with one particle in total on the three sites,  $\frac{1}{\sqrt{3}}\sum_{l=1}^3a_{j,l}^{\dag}e^{i\frac{2\pi}{3}(l-1)(k-1)}\ket{\textrm{vac}}$ for $k=1,2,3$ (which are eigenstates with eigenvalue $e^{i\frac{2\pi}{3} (k-1)}$) are mapped onto the three states $a_{j,k}^{\dag}\ket{\textrm{vac}}$. This is the same as in the general scheme, as there is just one particle, which consequently does not feel the hardcore constraint. From the particle-hole symmetry we deduce that the remaining three basis states $\frac{1}{\sqrt{3}}\sum_{l=1}^3a_{j,l}e^{i\frac{2\pi}{3}(l-1)(k-1)}a_{j,1}^{\dag}a_{j,2}^{\dag}a_{j,3}^{\dag}\ket{\textrm{vac}}$, $k=1,2,3$ (with eigenvalue $e^{i\frac{2\pi}{3} (k-1)}$), which have one hole, are mapped into the states that have the hole in copy $k$,  $a_{j,k}a_{j,1}^{\dag}a_{j,2}^{\dag}a_{j,3}^{\dag}\ket{\textrm{vac}}$. Clearly all these states can be distinguished in a measurement of the local occupation number. Note that, in contrast to the above general scheme where one requires the ability to measure the onsite particle number modulo three, this needs only a measurement modulo two. 
\section{Analysis of timing errors in the beamsplitter operation}

Here we analyse errors in the beamsplitter operation in step (2) of the measurement process. We assume that the ``rotation angle'' $\alpha$ is not perfectly chosen: $\alpha=\pi/4+\epsilon$, $\epsilon\ll1$, resulting in a operation \begin{align} a_{j,1}&\rightarrow \lr{1-\frac{\epsilon^2}{2}}\frac{a_{j,1}+a_{j,2}}{\sqrt{2}}+\epsilon \frac{a_{j,1}-a_{j,2}}{\sqrt{2}}, \\
a_{j,2}&\rightarrow \lr{1-\frac{\epsilon^2}{2}}\frac{a_{j,1}-a_{j,2}}{\sqrt{2}}-\epsilon \frac{a_{j,1}+a_{j,2}}{\sqrt{2}}.\end{align}
The measured expectation value is then \begin{align}\tr{V_2^{\mathcal{R}}(\epsilon)\rho\otimes \rho}\approx \tr{\rho_{\mathcal{R}}^2}+\epsilon^2 \tr{\mathcal{O}^{\mathcal{R}}V_2^{\mathcal{R}}\rho\otimes\rho}\end{align}
Note that, since $V_2^{\mathcal{R}}\rho\otimes\rho$ is a positive operator and $\mathcal{O}^{\mathcal{R}}=2\lr{\sum_{j\in \mathcal{R}}(a_{j,1}^{\dag}a_{j,2}-a_{j,2}^{\dag}a_{j,1})}^2$ is a negative one, these errors always lead to a measured value of $\tr{\rho_{\mathcal{R}}^2}$ that is  smaller than the actual one.

\end{document}